\documentstyle{elsart}

\def\degr{\hbox{$^\circ$}}
\def\lesssim{\mathrel{\hbox{\rlap{\hbox{\lower4pt\hbox{$\sim$}}}\hbox{$<$}}}}
\def\gtrsim{\mathrel{\hbox{\rlap{\hbox{\lower4pt\hbox{$\sim$}}}\hbox{$>$}}}}
\let\ga=\gtrsim
\let\la=\lesssim
\def\sun{\hbox{$\odot$}}

\input epsf.sty

\newcommand{\rmn}[1] {{\rm #1}}

\newcommand{\be}{\begin{equation}}
\newcommand{\ee}{\end{equation}}
\newcommand{\ba}{\begin{eqnarray}}
\newcommand{\ea}{\end{eqnarray}}
\newcommand{\brr}{\begin{array}}
\newcommand{\err}{\end{array}}
\newcommand{\bc}{\begin{center}}
\newcommand{\ec}{\end{center}}

\newcommand{\hMpc}{\mbox{$h^{-1}{\rmn{Mpc}}$}}
\newcommand{\hMpcI}{\mbox{$h\,{\rmn{Mpc}}^{-1}$}}

\begin{document}
\begin{frontmatter}
\title{Constraining cosmological models with cluster power spectra}
\author[AIP,MPE]{J. Retzlaff},
\author[PG]{S. Borgani},
\author[AIP]{S. Gottl\"ober},
\author[NMSU]{A. Klypin}, and
\author[AIP]{V. M\"uller}
\address[AIP]{Astrophysikalisches Institut Potsdam, 
     An der Sternwarte 16, D--14482 Potsdam, Germany}
\address[MPE]{{\em present address:\/} Max--Planck--Institut f\"ur
extraterrestrische Physik, Giessenbachstra{\ss}e, D--85740 Garching, Germany}
\address[PG]{INFN Sezione di Perugia, c/o Dipartimento di 
Fisica dell'Universit\`{a},
via A. Pascoli, I--06100 Perugia, Italy}
\address[NMSU]{Department of Astronomy, New Mexico State University,
    Las Cruces, NM 88001, USA}

\begin{abstract} 

  Using extensive $N$--body simulations we estimate redshift space
  power spectra of clusters of galaxies for different cosmological
  models (SCDM, TCDM, CHDM, $\Lambda$CDM, OCDM, BSI, $\tau$CDM) and
  compare the results with observational data for Abell--ACO clusters.
  Our mock samples of galaxy clusters have the same geometry and
  selection functions as the observational sample which contains 417
  clusters of galaxies in a double cone of galactic latitude
  $|b|>30\degr$ up to a depth of 240 $\hMpc$.
  
  The power spectrum has been estimated for wave numbers $k$ in the
  range $0.03 \la k \la 0.2\,\hMpcI$. For $k>k_{\rmn{max}}\simeq
  0.05\,\hMpcI$ the power spectrum of the Abell--ACO clusters has a
  power--law shape, $P(k)\propto k^n$, with $n \approx -1.9$, while it
  changes sharply to a positive slope at $k<k_{\rmn{max}}$. By
  comparison with the mock catalogues SCDM, TCDM ($n=0.9$), and also
  OCDM with $\Omega_0=0.35$ are rejected.  Better agreement with
  observation can be found for the $\Lambda$CDM model with
  $\Omega_0=0.35$ and $h=0.7$ and the CHDM model with two degenerate
  neutrinos and $\Omega_{\rmn{HDM}}=0.2$ as well as for a CDM model
  with broken scale invariance (BSI) and the $\tau$CDM model.
  As for the peak in the Abell--ACO cluster power spectrum, we find
  that it does not represent a very unusual finding within the set of
  mock samples extracted from our simulations.
\end{abstract}

\end{frontmatter}

\section{Introduction}

It is widely believed that cosmic large--scale structures have been
formed by gravitational instability from initially tiny Gaussian
density perturbations, which result from quantum fluctuations at very
early stages of cosmic evolution.  The Standard Cold Dark Matter
(SCDM) model is based on an inflationary model which predicts a
scale--invariant spectrum of perturbations (the Harrison--Zeldovich
spectrum) in an Einstein--de Sitter Universe.  Most of the matter
density in the Universe is provided by particles which interact only
gravitationally and which were non--relativistic when the Universe
became matter dominated. The nature of these dark matter particles is
still unknown. Candidates for the cold dark matter particles are both
the axions (light spin--0 bosons) and the lightest supersymmetric
particles which have masses over 100 GeV, while massive neutrinos are
candidates for hot dark matter particles.

However, the SCDM model is not compatible with existing observational data.
Once suitably normalized to reproduce clustering features at scales of
few Mpc, SCDM contradicts data at scales of few tens of Mpc, as well
as the CMB anisotropies detected by the {\sl COBE} satellite.  This
finding led to a large number of new theoretical models, the basic aim
of which was to predict less power than the SCDM model on small
scales, retaining, however, the enough power on the very large scales
probed by {\sl COBE}.

According to the standard picture of structure formation, the dark
matter power spectrum $P(k)$ is expected to bend from its
post--inflationary profile, and reaches a maximum at a wavenumber $k$
which depends on the parameters of the cosmological model. The
corresponding length scale is of the order of a few 100
$\hMpc$\footnote{$h$ is the Hubble constant in units of 100 km
s$^{-1}$ Mpc$^{-1}$}. One of the major aims of current cosmological
studies is to improve our knowledge of the structure of the
Universe on scales which are between those probed by the best 
redshift surveys of galaxies and those probed by {\sl COBE}.

Galaxy clusters are the largest gravitationally bound systems which we
observe. They arise from high peaks of the initial density field
(e.g., Bardeen et al.\ 1986) and have decoupled from the Hubble
expansion relatively recently. Therefore, their properties and their
spatial distribution are rather sensitive to the initial conditions
for the development of gravitational instability. Large cluster
samples include objects
up to a depth of a few hundred Mpc and, to date, 
they cover much larger volumes than any available galaxy redshift
survey.  For this reason much effort has been devoted to compile large
cluster samples, starting from the work by Abell (1958) and Abell,
Corwin \& Olowin (1989), and leading up to large redshift surveys both
in the optical (e.g., Postman, Huchra \& Geller 1992; Dalton et
al. 1994; Collins et al.\ 1995) and in the X--ray (e.g., Nichol, Briel
\& Henry 1994; Romer et al.\ 1994; Ebeling et al.\ 1997).  Employing
X--ray selected as well as optically selected catalogues from machine
based material is an attempt to overcome any subjective influence on
the selection criteria that is possibly present in the Abell
catalogue.

Historically, the analysis of the 2--point correlation function 
played an important role in
analyzing large--scale structure. After having recognized the
advantages of direct power spectrum estimation, this method has been
developed as a standard tool in cosmology. Nowadays, much valuable
information about structure formation is gained from power spectra of
galaxies in redshift surveys (e.g., Feldman, Kaiser \& Peacock 1994;
Park et al.\ 1994; Lin et al.\ 1996; Schuecker, Ott \& Seitter 1996;
Tadros \& Efstathiou 1996; cf.\ also Strauss \& Willick 1995, and
references therein).

Starting from the pioneering papers of Bahcall \& Soneira (1983) and
Klypin \& Kopylov (1983), who estimated the 2--point cluster
correlation function, galaxy clusters have been widely used for the
determination of parameters of large--scale structure. The power
spectrum of Abell clusters was calculated firstly by Peacock \& West
(1992), and, independently, by Einasto et al.\ (1993) and Jing \&
Valdarnini (1993).  Einasto et al.\ (1997a) have determined the
cluster power spectrum by inverting the 2--point correlation function of
the cluster redshift compilation by Andernach et al.\ (1995).

In this paper we compare the clustering properties of a redshift
catalogue of Abell--ACO clusters defined by Borgani et al.\ (1995) (cf.\
also Plionis \& Valdarnini 1991) with that of mock cluster samples
extracted from a set of $N$--body simulations of seven cosmological
models. Recently the same catalogue of Abell--ACO clusters has been
used to derive the Minkowski functionals (Kerscher et. al. 1997) of
the cluster distribution and to compare it with that of mock samples
from simulations.

This paper is organized as follows. In Sect.~\ref{sect:models} we introduce briefly
the cosmological models which we intend to discuss.  In Sect.~\ref{sect:clsample} we
describe the observational sample and the construction of mock samples
from numerical simulations. In Sect.~\ref{sect:pk} (and in a more technical
appendix) we describe the analysis of the cluster power spectrum.  In
Sect.~\ref{sect:results} we present and discuss our results.  We summarize and draw the
main conclusions in Sect.~\ref{sect:concl}.

\section{Cosmological models and numerical simulations}\label{sect:models}
 
Besides the SCDM model we consider six other models, which are
described in Table~\ref{tab:models}.  We used the transfer function by
Bardeen et al.\ (1986) and normalized the linear DM power spectra
according to either the two year {\it{COBE}} data (Bennett et al.\ 
1994, B94) or the four year {\it{COBE}} data (G\'orski et al.\ 1996,
G96; G\'orski et al.\ 1998, G98). The $\tau$CDM model is normalized to
the observed cluster abundance (Viana \& Liddle 1996, V96). We have
checked that a 20\% change in the normalization does not change the
power spectrum of the galaxy clusters selected from the simulations
once the number of clusters is fixed. In particular, for a given model
the result did not depend on the {\it COBE} normalization chosen
(first, second, or fourth year). This confirms earlier results by
Croft \& Efstathiou (1994) and Borgani et al.\ (1995). For this
reason, we also did not take into account any possible gravitational
wave contribution, which would reduce the BSI and TCDM normalizations.
We also did not take into account any correction to the power spectrum
shape due to a non--negligible baryon fraction.

The SCDM model with Harrison--Zeldovich primordial spectrum (as
predicted by the simplest inflationary models) assumes $\Omega_0=1$
and $h=0.5$. We consider it here as a reference model which, however,
is generally accepted to be ruled out, since it has a too shallow
power spectrum shape on intermediate scales ($10-50\,\hMpc$), and too
much power on small scales ($<10\,\hMpc$) once normalized to the
detected level of CMB anisotropy. 

An exponential inflaton potential leads to power law inflation and,
consequently, to a scale--free tilted (TCDM) model (e.g., Lucchin \&
Matarrese 1985). We assume here a spectral index $n=0.9$ after
inflation. As the result of the tilt the TCDM model with $h=0.5$ has a
lower normalization than SCDM in terms of the r.~m.~s.\ mass fluctuation
$\sigma_8$.

The CHDM model assumes a mixture of cold and hot dark matter. 
The 20\% contribution of the hot component is assumed to be shared 
between two neutrinos of equal mass (Primack et al.\ 1995). 

Lowering the matter content ($\Omega_0<1$) shifts the maximum of the
fluctuation spectrum to larger scales and steepens the spectrum 
on scales $\la 50\,\hMpc$.  In the $\Lambda$CDM model a cosmological
constant $\Omega_\Lambda \equiv \Lambda/(3H_0^2) = 1 - \Omega_0$ makes
the spatial curvature of the Universe negligible, as expected from
standard inflationary models. Our $\Lambda$CDM model assumes $h=0.7$,
a cosmological constant equivalent to $\Omega_\Lambda=0.65$.

The Open--bubble inflation model proposed by Ratra and Peebles (1994)
and the open model with scale--invariant spectrum (Wilson 1983) are
reasonably consistent with current observational data if $0.3 \la
\Omega_0 \la 0.6$ (G\'orski et al.\ 1998). We investigated the
Open--bubble inflation CDM (OCDM) model with $h=0.65$,
$\Omega_0=0.35$ and the normalization by G98.

The Broken Scale Invariant (BSI) cosmological model is specified by
two parameters, the step location at $k_{\rm{break}}^{-1}=1.5\,\hMpc$
and its relative height $\Delta=3$; the normalization is $\sigma_8 =
0.60$. These parameters are related to the initial energy densities
and mass ratios of the inflaton fields in the underlying inflationary
model (Gottl\"ober, M\"uller \& Starobinsky 1991, Gottl\"ober 1996).
The original choice for such parameters has been based on the
linear--theory comparison with different observational constraints
(Gottl\"ober, M\"ucket \& Starobinsky 1994). Analysis based on $N$--body
simulations have been discussed by Amendola et al.\ (1995), Kates et
al. (1995) and Ghigna et al.\ (1996). Recently, Lesgourgues et al.\
(1997) have discussed a BSI spectrum in a model with cosmological
constant.

The $\tau$CDM model, which assumes a decaying massive $\tau$ neutrino
as the dark matter constituent, has been originally proposed by
Efstathiou et al.\ (1992a; see also White et al.\ 1995). Similarly to
the BSI model, the $\tau$CDM model is characterized by two additional
free parameters, namely the mass and the life time of the decaying
massive particle, which are related to the scale where the spectrum
changes and the amount of small--scale power suppression relative to
the SCDM case.  We consider the CDM model with a shape parameter of
$\Gamma=0.21$ as a representation of the $\tau$CDM model (Jenkins et
al.\ 1998).  Normalized to the observed abundance of clusters (V96)
this spectrum is very similar to the {\it COBE} normalized BSI
spectrum.

\begin{table}[tp]
\caption{The model parameters.
Column 2: the density parameter $\Omega_0$; 
Column 3: the Hubble parameter $h$; Column 4: the  {\sl COBE} predicted
linear r.~m.~s.\ fluctuation amplitude at $8\hMpc$ $\sigma_8$; Column 5: 
reference to the normalization, Column 6: further model parameters}
\medskip
\tabcolsep 5pt
\begin{tabular}{lcccll} \hline
 Model & $\Omega_0$  & $h$  & $\sigma_{8}$ & normalization & model parameters \\ \hline
SCDM          & 1.0  & 0.50 & 1.37 &  B94 &\\
TCDM          & 1.0  & 0.50 & 1.25 &  B94 & $n$ = 0.9 \\
CHDM          & 1.0  & 0.50 & 0.67 &  G96 & $\Omega_{\rmn{HDM}} = 0.2$, 2$\nu$ \\ 
$\Lambda$CDM  & 0.35 & 0.70 & 1.30 &  B94 & $\Omega_{\Lambda} = 0.65$ \\
BSI           & 1.0  & 0.50 & 0.60 &  B94 &
$k_{\rm{break}}^{-1}=1.5\,\hMpc$,  $\Delta=3$\\
$\tau$CDM     & 1.0  & 0.50 & 0.60 &  V96 & $\Gamma = 0.21 $ \\
OCDM          & 0.4  & 0.65 & 0.58 &  G98 & \\ \hline
\end{tabular}
\label{tab:models}
\end{table}

We evolve the initial density field starting from redshift $z=25$
($z=30$ for CHDM) until the present epoch, by employing a PM $N$--body
code with $300^3$ particles of mass $m_{\rmn{p}} = 1.3
\times 10^{12}\, h^{-1}\Omega_0{\rmn{M_{\sun}}}$. Cold particles in
the CHDM simulations have a mass which is 20\% smaller than this
value. The simulation used $600^3$ grid cells in a
simulation box of $L=500\,\hMpc$ comoving length a side.  This
provides a formal spatial resolution of less than $1 \,\hMpc$.  The
simulation box is large enough to contain all fluctuation modes which
contribute to the large--scale cluster distribution at the scales we
are interested in.  In order to account for the effect of cosmic
variance, we carried out simulations for several random realizations
(four for SCDM, three for $\Lambda$CDM, eight for CHDM). For the
remaining models we did only one realization because the effect of
cosmic variance appears to be similar as for the other models under
consideration.


\section{Cluster samples}\label{sect:clsample}
\subsection{The observational sample}\label{sect:obs}

The sample that we will consider includes Abell and ACO clusters with
richness $R\ge 0$ (Abell 1958; Abell, Corwin \& Olowin 1989).  Here we
will provide a brief description of this sample which was first
defined in Plionis \& Valdarnini (1991) and updated by Plionis \&
Valdarnini (1995) and Borgani et al.\ (1997).  Initially, the northern
(Abell) part of the sample, with declination $\delta\ge -17\degr$, was
defined by those clusters that have measured redshift $z\le 0.1$,
while the southern ACO part, with $\delta < -17\degr$, was defined by
those clusters with $m_{10}< 17$, where $m_{10}$ is the magnitude of
the tenth brightest cluster galaxy.  In our analysis we include only
clusters with redshifts smaller than 0.085. We checked that in that
case the redshift limit adopted for the Abell part and the $m_{10}$
limit applied to the ACO part are essentially equivalent, the samples
are 97\% complete in this range.

The galactic absorption is modeled according to the
standard cosecant dependence on the galactic latitude $b$,
\begin{equation} \label{eq:obs1}
\varphi(b)=10^{\alpha\left(1-\csc|b|\right)} \; ,
\end{equation}
with $\alpha = 0.3$ and 0.2 for Abell and ACO parts of the sample,
respectively. In order to limit the effects of galactic absorption we
only use clusters with $|b|\ge 30\degr$.

\begin{figure}
\epsffile{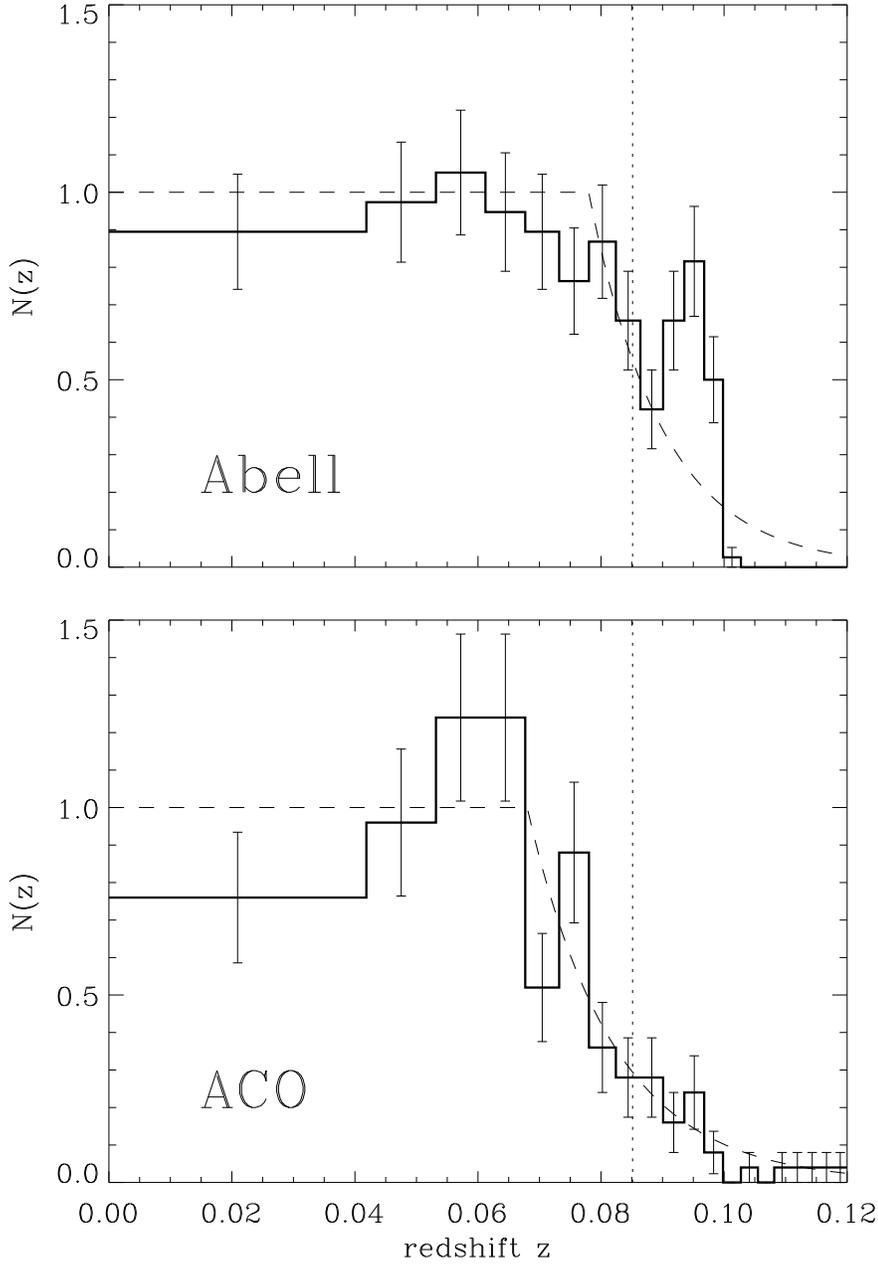}
\caption{Normalized redshift distribution of clusters, estimated in
equa--volume shells, for the Abell sample (upper panel) and ACO sample
(lower panel). The fits with exponential tails and parameters as given
in the text are shown with dashed lines. The dotted vertical line
indicates the adopted limiting redshift.}
\label{fig-dz2}
\end{figure}

The cluster--redshift selection function, $\psi(z)$, is determined by
fitting the cluster density as a function of~$z$,
\begin{equation} \label{eq:pz}
\psi(z)=\left\{ \begin{array}{ll}
 1               & \mbox{if $z\le z_{\rmn{c}}$} \\
 A\,\exp(-z/z_{\rmn{o}}) & \mbox{if $z>z_{\rmn{c}}$}
\end{array}
\right. \; ,
\end{equation} 
where $A=\exp{(z_{\rmn{c}}/z_{\rmn{o}})}$, and $z_{\rmn{c}}$ is the
redshift below which the spatial density of clusters remains constant
and the sample behaves as a volume--limited one. The best--fitting
values for such parameters are $z_{\rmn{c}} = 0.078$, $z_{\rmn{o}} =
0.012$ and $z_{\rmn{c}} = 0.068$, $z_{\rmn{o}} = 0.014$ for Abell and
ACO samples.  We convert redshift into distance using the Mattig
formula with $q_0=0.5$ for the deceleration parameter. We also checked
that final results of the power spectrum analysis remain unchanged
taking instead $q_0=0.2$. Since the exponential decrease of $\psi(z)$
introduces considerable shot noise errors at large redshifts, we
prefer to limit our analysis to $z=0.085$, which corresponds to
$r_{\rmn{max}}=240\,\hMpc$ for $q_0=0.5$.  Fig.~\ref{fig-dz2}
illustrates the fit of the exponential tail to the redshift
distribution for Abell and ACO samples. The dotted vertical lines
indicate the adopted limiting redshift.

There are in total 417 clusters fulfilling the above criteria: 262
Abell clusters with measured redshifts, and 155 ACO clusters,
139 with measured redshifts and 16 with redshift estimates from the
$m_{10}$-$z$ relation defined in Plionis and Valdarnini (1991).  This
corresponds to $\langle n \rangle_{\rmn{Abell}} \simeq 1.8 \times
10^{-5}\,h^3{\rmn{Mpc}}^{-3}$ and $\langle n \rangle_{\rmn{ACO}}
\simeq 2.5 \times 10^{-5}\,h^3{\rmn{Mpc}}^{-3}$, for the Abell and ACO
cluster number densities, respectively, once corrected for galactic
absorption and radial selection according to Eqs.\ (\ref{eq:obs1}) and
(\ref{eq:pz}). The above density values correspond to average cluster
separations of $\langle d \rangle_{\rmn{Abell}} \simeq 38\,\hMpc$ and
$\langle d \rangle_{\rmn{ACO}} \simeq 34\,\hMpc$.  This density
difference has been shown to be mostly spurious, due to the higher
sensitivity of the \mbox{IIIa--J} emulsion plates on which the ACO
sample is based (see, e.g., Batuski et al.\ 1989; Scaramella et al.\
1991; Plionis \& Valdarnini 1991).  It is important to account
properly for the difference in density in order to avoid spurious
large--scale power in the analysis.  We use the ratio of densities
($D\equiv\langle n \rangle_{\rmn{Abell}} / \langle n
\rangle_{\rmn{ACO}} \simeq 0.7$) as an overall weighting factor for
the Abell part.

The reliability of the Abell--ACO sample for clustering analysis has
been questioned by several authors (e.g., Sutherland 1988; Efstathiou
et al.\ 1992b). They argued that part of the strong clustering
exhibited by Abell--ACO clusters is due to spurious projection
contamination, which enhances the correlation amplitude along the line
of sight. Although such an effect is undoubtedly present in the
Abell--ACO sample, the amount of contamination it introduces in
clustering measurements is matter of debate. For instance, Jing,
Plionis \& Valdarnini (1992) analyzed cluster simulations in order to
check the significance of the enhanced line--of--sight correlation
amplitude measured for real data. They found that anisotropy in the
correlation function as large as the observed one is rather common in
simulations and is generated by statistical fluctuations. Olivier et
al.\ (1993) claimed that any contamination in the Abell--ACO sample
should in any case have a negligible effect on the 2--point
correlation function $\xi(r)$, on scales larger than superclusters,
i.e. $\gtrsim 20-25\,h^{-1}$Mpc.

A further potential problem in the analysis of the distribution of
Abell--ACO clusters is associated with the obscuration in our Galaxy.
The patchy distribution of gas could introduce a spurious modulation
in the cluster distribution, which can not be fully corrected by
resorting to the simple $\csc|b|$ relation of
Eq.~(\ref{eq:obs1}). Nichol \& Connolly (1996) studied the
cross--correlation between the angular distribution of Abell clusters
and galactic HI measurements. They concluded that there is a
statistically significant anticorrelation between the distributions of
HI regions and Abell clusters having higher richness and distance
class. On the other hand, nearby $R=0$ clusters appear to be randomly
distributed with respect to the galactic HI column density. 

In order to verify whether such potential biases in the Abell--ACO
sample quantitatively affect clustering measures, we compute the
cluster 2--point correlation function, $\xi(r)$, and compare it with
recent results for the APM cluster sample (Croft et al.\ 1997), which
is in principle much less affected by projection contamination and
patchy obscuration. In order to estimate $\xi(r)$, we resorted to 
the estimator
\be
\xi(r)\,=\,{DD(r)\over DR(r)}-1\,.
\label{eq:xi}
\ee 
Here $DD(r)$ and $DR(r)$ are the number of data--data and
data--random cluster pairs at separation $r$. The quantity $DR$ is
estimated by averaging over 200 random samples, each having the same
redshift selection function and galactic extinction as the real
catalogue.

The result of this analysis is reported in Fig.~\ref{xir}. The error
bars correspond to the 1$\sigma$ scatter between the $\Lambda$CDM mock
samples (see below). We resort to a log--log weighted least--square
fit to estimate the correlation length $r_0$ and the slope $\gamma$
for the power--law model, $\xi(r)=(r_0/r)^\gamma$. Assuming this model
over the scale range $4\la r\la 50\,h^{-1}$Mpc, we obtain $r_0=(16.7
\pm 3.2) h^{-1}$ Mpc and $\gamma = 2.15 \pm 0.15$ (the power--law
corresponding to the best fitting parameters is plotted as a dashed
line in Fig.~\ref{xir}). Our result is very close to that obtained by
Croft et al.\ (1997) over a similar scale range, for APM subsamples
having comparable cluster average separation. Indeed, they found
$r_0=(14.2\pm 0.5)\,h^{-1}$Mpc and $\gamma =2.13\pm 0.08$
($r_0=(16.6\pm 1.3)\,h^{-1}$Mpc and $\gamma =2.1\pm 0.1$) for an APM
subsample with mean cluster separation of $d_{cl}=30\,h^{-1}$Mpc
($d_{cl}=48\,h^{-1}$Mpc). This indicates that any bias in the
Abell--ACO sample is not so effective as to heavily pollute the
estimate of the cluster 2--point correlation function.  One can not
strictly infer the behavior of the power spectrum from the correlation
function since the sensitivity to biases is different for the two
statistics.  However, we consider the result from the correlation
function at least as an indication that the power spectrum estimation
is not dominated by a possible bias.

\begin{figure}
\epsfxsize \textwidth
\epsffile{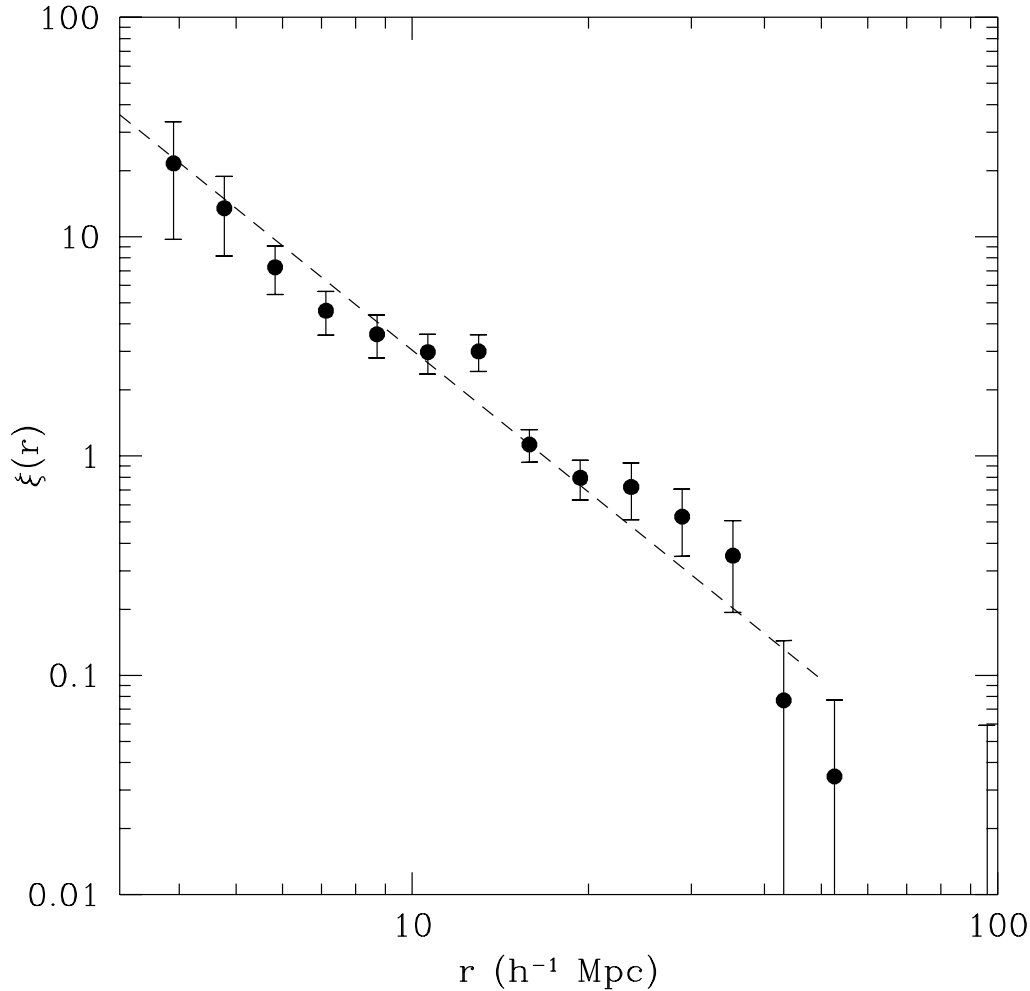}
\caption{
Redshift space cluster 2--point correlation function for the
Abell--ACO sample. The error bars corrspond to the 1$\sigma$ cosmic
scatter, as derived from numerical
simulations. The dashed line represents the best--fitting power--law
model for $\xi(r)$.}
\label{xir}
\end{figure}


\subsection{Mock samples in simulations}

We have used two different methods to identify galaxy clusters in
the simulations. The first is based on the friend--of--friend (FOF)
algorithm and the second on the peak--in--density algorithm (Klypin \&
Rhee 1994). As a first step, we found either the centers of mass
for the FOF groups or the location of the grid points corresponding
to local density maxima. Then we centered spheres with Abell
radius 1.5 $h^{-1}$ Mpc around these points and determined the centers
of mass of the particles falling into such spheres. Using the centers
of mass as new cluster centers after few iterations the position of the
centers rapidly converge. The cluster distributions obtained by
starting from the FOF points and from the density peaks turned out to
be virtually identical, not only in a statistical sense, but also
through a point--by--point comparison (see also Klypin et al.\ 1997,
where similar halo finding algorithms have been discussed).  From the
resulting list of candidate clusters, we select the $N_{\rmn{c}}$ most
massive objects to be identified as Abell--ACO clusters. By definition,
$N_{\rmn{c}}=(L/\langle d \rangle)^3$ is the expected number of
clusters within the simulation box, where $\langle d \rangle
=34\,\hMpc$ is the average separation appropriate for Abell--ACO
clusters.

After generating the cluster distribution in the simulation box, we
extract mock samples with the same geometrical boundaries and
selection functions as for the real sample. In each box we locate 8
observers along the main diagonal axes, each having a distance of
$L/4=125\,\hMpc$ from the three closest faces.  For each observer we
construct the redshift space distribution of clusters starting from
the real space positions and peculiar velocities.  First we include
all clusters up to a maximum redshift $z=0.085$ in each mock sample,
which corresponds to a distance of $240\,\hMpc$ for $q_0=0.5$.  Then,
we randomly sample the cluster population so as to obtain a density
distribution which reproduces the observational selection functions
for galactic absorption and redshift extinction.  Finally, we randomly
remove 30\% of the cluster in the Abell part to get the observed
relative density with respect to the ACO part as in the real sample.
This procedure of random dilution in the Abell portion of the mock
samples is consistent with the expectation that real clusters are
missed in the observational sample, due to the lower sensitivity of
the Abell emulsion plates with respect to those (IIIa--J) from which
the ACO clusters have been identified.  It is clear that our procedure
to account for the different density in the two parts of the sample
would not be correct in case the Abell cluster selection picks up
intrinsically richer system than the ACO one. However, even in this
case, the 30\% difference in the cluster number density would
correspond to only a $\sim 10\%$ difference in their mean separation.
Therefore, any realistic clustering--richness relation would only
induce a marginal difference between the correlation amplitudes of
Abell and ACO clusters.

In order to minimize the overlap between mock samples,
the coordinate systems for two adjacent observers are chosen so that
the corresponding galactic planes are orthogonal to each other. Even
with this choice, it turns out that different mock samples involve
small overlapping volumes and, therefore, they cannot be considered as
completely independent.

Various effects may give rise to systematic errors in the measurement
of the power spectrum. In order to quantify these effects, that we
describe here below, we employed Monte Carlo techniques.  Very nearby
clusters of galaxies are hard to detect on the sky because of the
large angular spread of their galaxy members. As a consequence, these
objects do not meet the Abell selection criteria and they are simply
missed from the sample. In order to account for this, we excluded from
the analysis of mock samples those clusters which are closer than a
fixed minimum distance. As a conservative choice, we fixed it to
$40\,\hMpc$ and verified that no change in the power spectrum
estimation occurs. This is not surprising since the corresponding
volume fraction with respect to the whole sample (of depth
$240\,\hMpc$) is less than one per cent.  We have also checked that a
change in the galactic extinction selection function from $\alpha=0.2$
to $0.3$ does not alter the power spectrum results.


\section{The power spectrum}\label{sect:pk}

The distribution of galaxy clusters is interpreted as a random point
process with the power spectrum being the first non--trivial
quantitative description in a hierarchy of statistical measures.  We
use an estimate of the power spectrum which is appropriate for the
available finite sample of clusters. The same procedure is applied to
Abell--ACO clusters (observational sample) as well as to the simulated
clusters (mock samples).

According to Eq.~(\ref{tildeP}) the power spectrum estimate
$\tilde{P}'(k)$ of the finite sample of galaxy clusters is given by the
convolution of the true power spectrum of the cluster distribution
with the window function of the cluster sample. According to
Eq.~(\ref{conv}), it is
\begin{eqnarray}
\lefteqn{ \tilde{P}'(k) = 
\frac{V}{1-\langle | \hat{W}(\vec{k})|^2\rangle_{|\vec{k}|=k}} \times} 
\nonumber \\
\lefteqn{ \left\{ \frac{1}{M}\sum_{j}\left|\frac{\sum_{i}\phi^{-1}(\vec{r}_i)
e^{i\vec{k}_j\cdot\vec{r}_i}}{\sum_{i}\phi^{-1}(\vec{r}_i)} -\hat{W}(\vec{k}_j)\right|^2 
- \frac{\sum_{i}\phi^{-2}(\vec{r}_i)}{\left(\sum_{i}\phi^{-1}(\vec{r}_i)\right)^2} \right\}\;,} \nonumber \\
{}\label{final-estimator} & & 
\end{eqnarray}
where $\hat{W}(\vec{k})$ is the Fourier representation of the window
function (i.e., the volume encompassed by the sample), $\phi(\vec{r})$
is the selection function which incorporates galactic extinction,
redshift selection and the density difference between Abell and ACO
parts of the sample.  The sum $\sum_{j}$ extends over $M$ random
directions for the wave vector $\vec{k}_j$, while $\sum_{i}$ extends
over the sample points. 

We do not attempt any deconvolution procedure of $\tilde{P}'(k)$ for
the following two reasons. Firstly, our main aim is the comparison of
observational results with numerical simulation outcomes. We generate
mock samples from the simulations so as to have the same properties
(in terms of geometry and selection functions) as the observational
data. This procedure ensures that we have the same effect due to
window convolution in the mock samples analysis. Secondly, we take
advantage of the availability of the parent cluster distributions
within the periodic simulation box, from which the mock samples are
extracted, to specify the $k$ range where the finite window does not
affect the $P(k)$ estimate.

\begin{table}
\caption{\label{tab-sample} 
Overview of the observational samples. Column 2:
total number of objects, $N$; column 3: completeness factor $C$:
column 4: volume $V$; column 5:
solid angle $\Omega$; column 6: shot noise power level $P_{\rmn{noise}}$.}
\medskip
\begin{tabular}{lrrrrr} \hline
Sample & $N$ & $C$ & $V$ & $\Omega$ & $P_{\rmn{noise}}$ \\
 & & & $(h^{-3}{\rmn{Mpc}}^3)$ & $(4\pi)$ & $(h^{-3}{\rmn{Mpc}}^3)$ \\ \hline
Abell--ACO & $417$ & $0.56$ & $2.9\times 10^7$ & $0.50$ & $7.7\times 10^4$ \\
Abell & $262$ & $0.74$ & $2.0\times 10^7$ & $0.34$ & $8.1\times 10^4$ \\
ACO & $155$ & $0.67$ & $9.2\times 10^6$ & $0.16$ & $6.9\times 10^4$ \\
ACO--SG & $136$ & $0.67$ & $8.0\times 10^6$ & $0.14$ & $6.9\times 10^4$ \\ \hline
\end{tabular}
\end{table}

In Table~\ref{tab-sample} we summarize the main global properties for
the Abell--ACO observational cluster sample, as well as for three
subsamples (Abell, ACO, and the southern--galactic portion of the ACO
sample, ACO--SG).  The statistical completeness factor, $C$, is
defined as the ratio of the actual number of objects and the expected
number if no selection effects were present.


\section{Results and Discussion}\label{sect:results}

\begin{figure}
\epsfxsize \textwidth
\epsffile{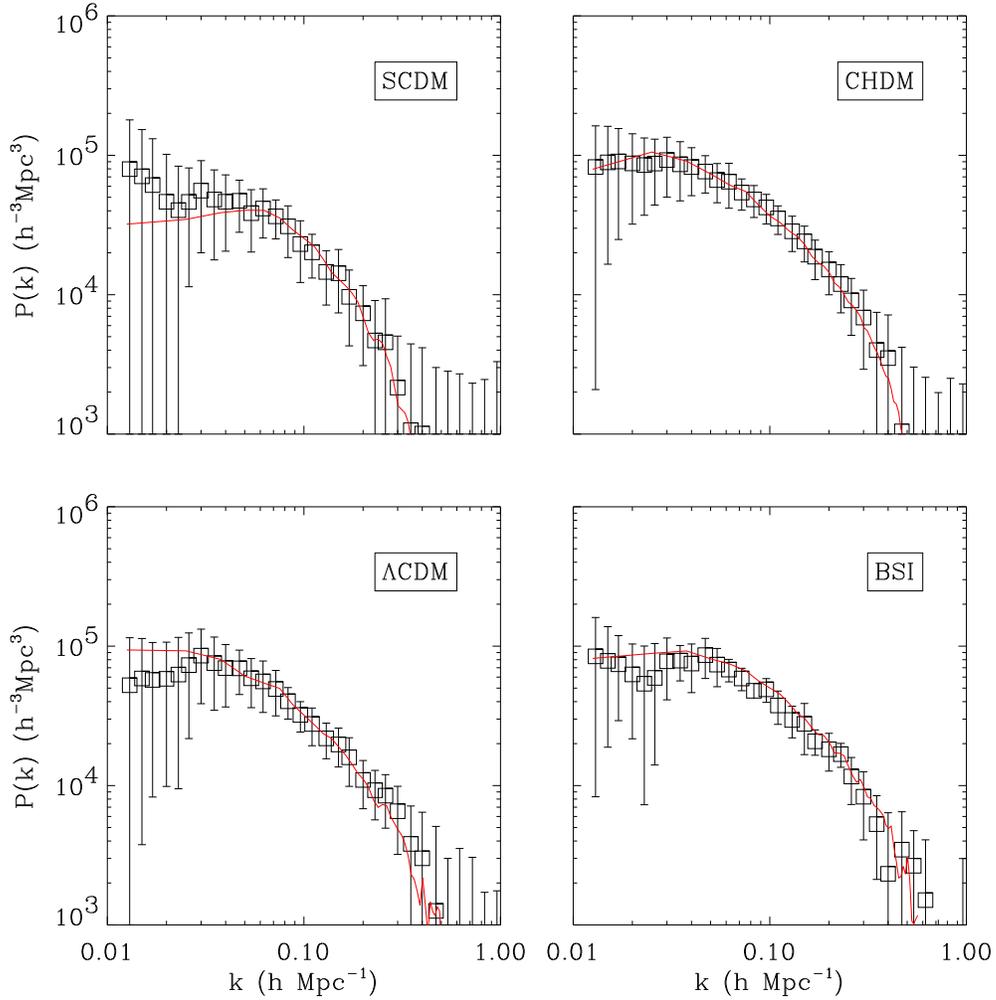}
\caption{
Comparison of the convolved power spectrum estimated from
mock samples (square symbols) with the `real' redshift space power spectrum from
whole--box simulated catalogues (solid line).}
\label{fig-mockbox}
\end{figure}

We have calculated the mean value and the standard deviation of the
power spectra within the set of mock samples available from the
$N$--body simulations for each of the cosmological model.  In order to
assess the reliability of our method to estimate the power spectrum, we
have compared the power spectrum $P(k)$ of simulated clusters in the
periodic simulation box with the power spectrum estimate
$\tilde{P}'(k)$ of the mock catalogues. 
For the clusters in the periodic simulation cube a real--to--redshift
space transformation along one dimension has been applied before.

In Fig.~\ref{fig-mockbox} the solid line denotes the mean $P(k)$
calculated from a number of simulations for the corresponding four
cosmological models. The squares with error bars denote the mean of
the convolution of the power spectrum $\tilde{P}'(k)$ as found from
the mock samples. First of all, this is a test for the correct
implementation of the method. Moreover it provides an insight about
the limitations of the method on the largest scales, where window
convolution effects limit the reliability of the analysis. We conclude
that the influences of the window on the reconstruction method is
negligible for $k > 0.03\,\hMpcI$. In the following, we omit the tilde
and prime on the convolution of the power spectrum with the window
function.

We find that the variance between power spectra of different
realizations is almost identical for different models. Therefore, we
assume in the following the resulting cosmic scatter to be
representative also for the observational results.  Though this method
closely resembles a Monte Carlo error estimation technique, we are
restricted here to a smaller number of random realizations than
usually employed. In the following we will take the $1\sigma$ scatter
between the set of 64 CHDM mock samples as the error to be assigned to
the Abell--ACO power spectrum since this model has the largest number
of available realizations.

\begin{figure}
\epsffile{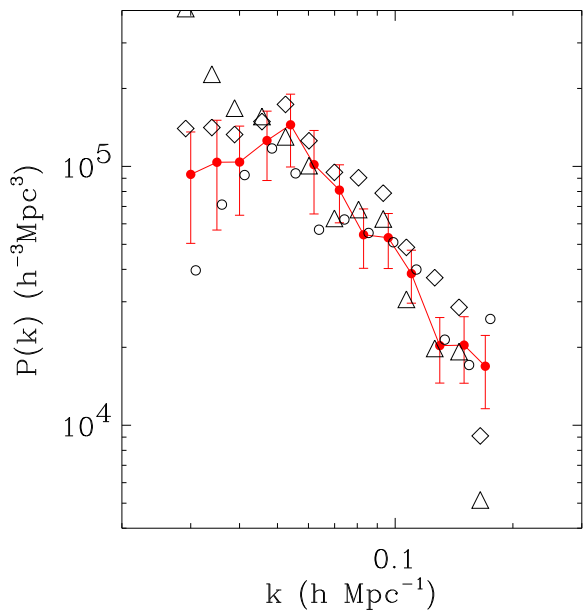}
\caption{
Redshift space cluster power spectrum for the Abell--ACO sample ---
filled circles, the Abell part --- open circles, and the ACO part ---
diamonds (only southern--galactic part --- triangles).  The 1$\sigma$
error bars are derived from numerical simulations.}
\label{fig-obs}
\end{figure}

We show in Fig.~\ref{fig-obs} the results of the power spectrum
reconstruction for the observational data.  Open symbols refer to the
different subsamples which have the same depth but cover different
regions on the sky. Filled circles are for the combined Abell--ACO
sample. For reasons of clarity, we plot error bars only for the
latter.  We find a general good agreement between the Abell and ACO
parts (open circles and diamonds, respectively), and the combined
sample, although ACO clusters show a slightly higher $P(k)$ on all
scales. This result, which is consistent with the larger correlation
length found for ACO clusters in previous analyses (e.g., Cappi \&
Maurogordato 1992; Plionis, Valdarnini \& Jing 1992), can be
attributed to a particular structure, the Shapley supercluster, which
is located in the northern galactic part of the ACO sample. Excluding
this region from the analysis (triangles) decreases the power spectrum
amplitude (ACO--SG), although still within the cosmic variance error
bars.  Since each single subsample covers a smaller volume than the
combined sample, finite volume effects start playing a role at smaller
$k$ values (cf.\ Fig.~\ref{fig-w2k}, where we plot the window function
shape for each subsample). Especially for ACO--SG, the amplitude at
small wavenumbers is blown up.  On scales $k > 0.05\,\hMpcI$ the
spectrum may be approximated by a power law, $P(k)\propto k^n$ with
negative index $n \simeq -1.9$.  Going to larger scales, there is a
clear evidence for a breaking of the power law.  The transition is
located at about $k \simeq 0.05\,\hMpcI$ beyond which a power law with
positive index could fit the data. Due to the large error bars, the
scale of transition can not be pinned down very accurately.

\begin{figure}
\epsffile{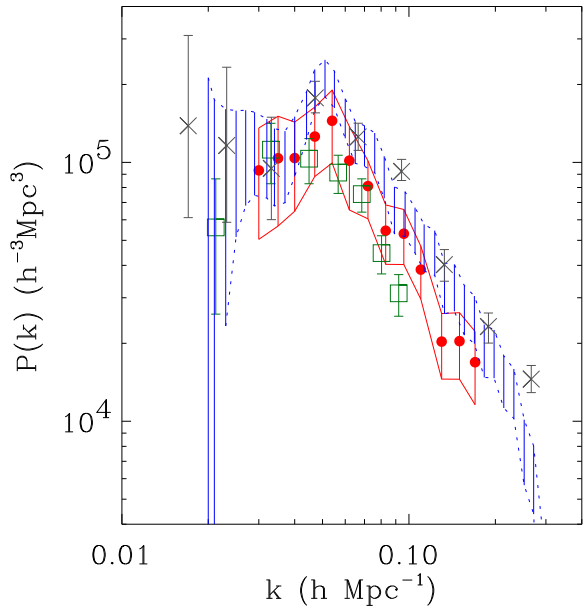}
\caption{
Comparison of various cluster power spectra obtained from 195 Abell clusters
by Peacock \& West (1992, crosses), 1304 Abell--ACO clusters by 
Einasto et al.\
(1997a, bars/dashed lines), 364 APM clusters by Tadros, Efstathiou \&
Dalton (1997, squares) and 417 Abell--ACO clusters (this work, dots and
solid lines).}
\label{obspk_new}
\end{figure}

\begin{figure}
\epsfxsize \textwidth
\epsffile{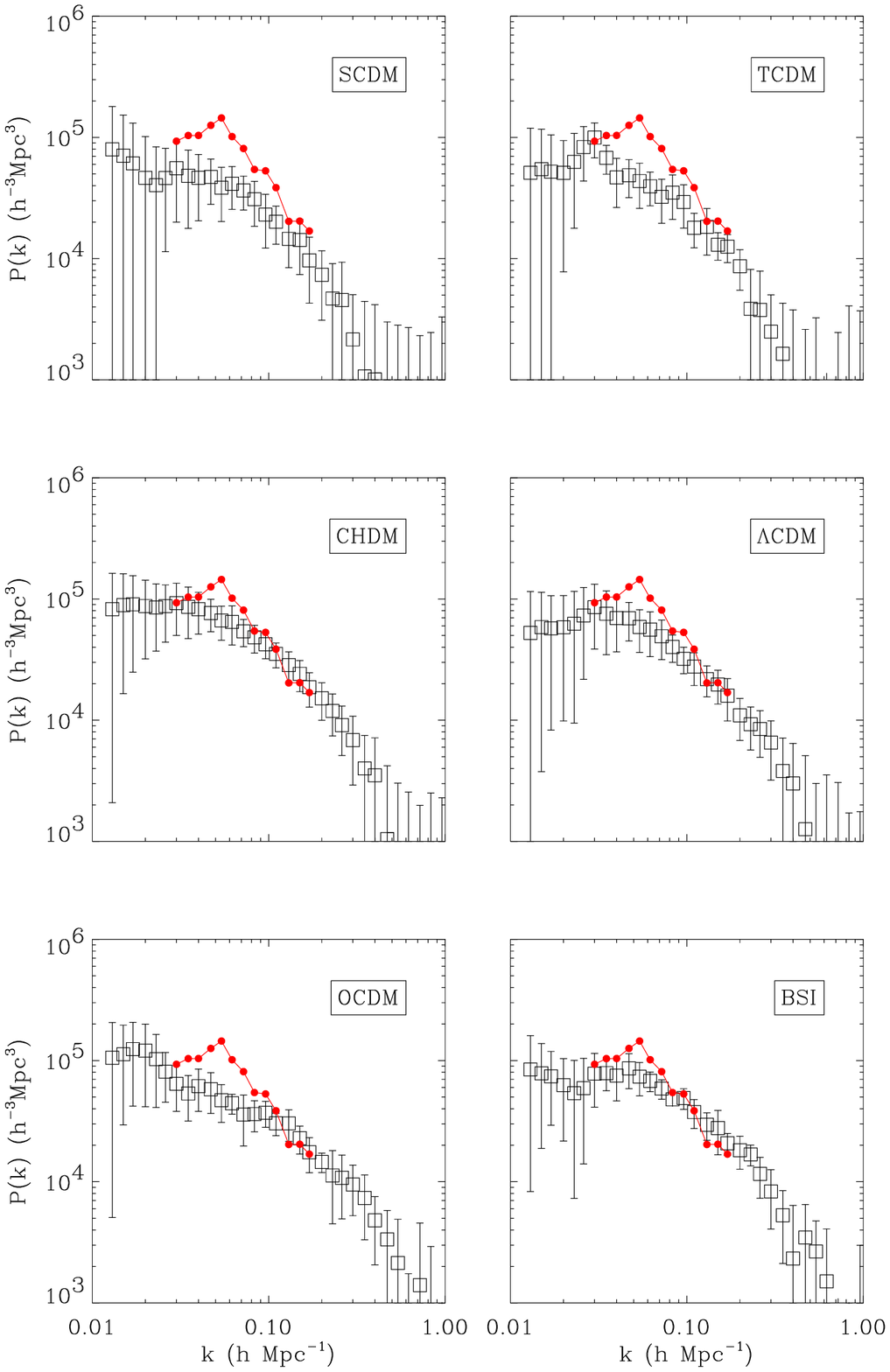}
\caption{
Power spectrum of the observed cluster sample (filled circles)
compared with results obtained from mock cluster samples (squares) for
six cosmological models.  Error bars denote the $1\sigma$ variation
over the mock samples.}
\label{fig-mock}
\end{figure}

Recently, the cluster power spectrum has been estimated for a sample
of 1304 Abell/ACO clusters of galaxies by Einasto et al.\ (1997a).
Differently from the analysis presented here, Einasto et al.\ (1997b)
determined the large--scale behavior of the correlation function, and
then used its Fourier transform as an estimate of the power spectrum.
This method leads to a power spectrum with a pronounced peak at about
$100 \hMpc$ scale (see Fig.~\ref{obspk_new}).  The error corridor
(indicated by dashed lines), which was also estimated from the
correlation function, turns out to be narrower than our cosmic
variance uncertainties.  The parent cluster catalog analyzed by
Einasto et al.\ (1997b) --- the Abell--ACO catalog --- is the same as
ours. However, although the selection of $R\ge 0$ clusters within the
angular boundaries dictated by galactic absorption is identical, the
depth is different. Indeed, the sample used by Einasto et al.\ extends
to a depth of 340 $h^{-1}{\rm{Mpc}}$ at the price of a higher fraction
of non--spectroscopical cluster redshifts (33\%), when compared to the
4\% fraction of the sample we used here.  Besides the difference in
the observational data, we use different methods for estimating the
power spectrum. In any case, the fact that the same $P(k)$ slope,
$n\approx -1.9$, is found with both methods is a support to the
robustness of this result. The maximum of $P(k)$ that we detect at
$k=0.054\,\hMpcI$ is somewhat less significant than, although
consistent with, that found by Einasto et al.  The significance of the
peak is further decreased once the cosmic scatter is taken into
account.  The higher $P(k)$ amplitude found by Einasto et al.\ should
be attributed to the different sample of 1304 clusters out to
$340\,\hMpc$, and to uncertainties in the normalization of the
2--point correlation function that was used as the starting point of
their analysis. In Fig.~\ref{obspk_new}, we also show the previous
results from Peacock \& West (1992) for $R \ge 1$ Abell clusters.  At
wave numbers $k > 0.08 \,\hMpcI$, the resulting $P(k)$ even tends to
lie above the estimate by Einasto et al., as it should be expected on
the ground of the higher richness of the clusters considered by
Peacock \& West (1992).  Also their estimated errors are much smaller
than ours, thus indicating the necessity to perform a large number of
$N$--body simulations for estimating statistical uncertainty and
cosmic variance.  The power spectrum of rich clusters selected from
the APM galaxy survey (Tadros, Efstathiou \& Dalton 1997) turns out to
be essentially consistent with our results.  For wave numbers $k >
0.04 \,\hMpcI$, the spectral amplitude is slightly smaller while the
spectral slope is absolutely compatible. This consistency confirms the
result based on the 2--point correlation function (cf.\ Fig.~\ref{xir}
and \S\ref{sect:obs}). It again indicates that Abell--ACO and APM
cluster samples provide statistically equivalent descriptions of the
local ($z\la 0.1$) cluster distribution.

In the following we will compare the power spectrum of the combined
Abell--ACO sample with simulation results.  In Fig.~\ref{fig-mock} the
results from the mock samples (squares) are compared with the
observational results (filled circles). Simulation results are the
average over the available mock samples and error bars are the
corresponding $1\sigma$ scatter between them.  Only for SCDM, CHDM and
$\Lambda$CDM error bars can be considered as a reliable representation
of cosmic variance, since four, eight and three independent numerical
realizations were carried out for these models, respectively.  The
predicted power spectrum of SCDM is too small on all scales. This
confirms that the model definitely underproduces cluster clustering.
Even by allowing for an overall vertical shift, the slope of the
predicted spectrum is too shallow on intermediate scales $k=0.06 -
0.2\,\hMpcI$. The TCDM model provides a nearly as worse fit as SCDM
does.  The CHDM model provides a good fit on small scales, but it
underpredicts the $P(k)$ amplitude of the observational sample around
the maximum, $ 0.04 \la k \la 0.07\,\hMpcI$.  Also the $\Lambda$CDM
model is able to reproduce the power on large scales ($k \simeq
0.035\,\hMpcI$) and small scales ($k \simeq 0.15\,\hMpcI$) quite well,
but it fails on intermediate scales having a too shallow slope.  The
OCDM model underpredicts cluster clustering by a similar amount as
SCDM and TCDM. As for BSI, since error bars are likely to be slightly
underestimated (just a single simulation realization is available), it
provides the best fit to the data among the models we considered.
However, even in this case, the overall shape of the observational
$P(k)$ for $0.055\la k \la 0.15\,\hMpcI$ is steeper than for BSI
clusters. We did not include the $\tau$CDM model into
Fig.~\ref{fig-mock} because it is quite similar to the BSI model. This
is not surprising since their linear spectra are almost identical.

In order to quantify the systematic discrepancy between the $P(k)$
shape for data and simulations, we performed a least--square fitting to
the power law $P(k)\propto k^n$ in the above $k$ range. As a result,
we find $n=-1.9\pm 0.2$ for the Abell--ACO sample, while $n=-1.3\pm
0.2$ (SCDM), $-1.2\pm 0.2$ (TCDM), $-1.3 \pm 0.2$ (CHDM), $-1.2\pm
0.2$ ($\Lambda$CDM), $-1.0 \pm 0.2 $ (OCDM), $-1.1\pm 0.2$ (BSI), and
$-1.1 \pm 0.2 $ ($\tau$CDM). At this level, we consider as premature to
decide whether such a $\sim 2\sigma$ discrepancy between real data and
mock samples is just due to a statistical fluctuation or is indicating
an intrinsic problem for standard DM power spectra.

\begin{figure}
\epsfxsize \textwidth
\epsffile{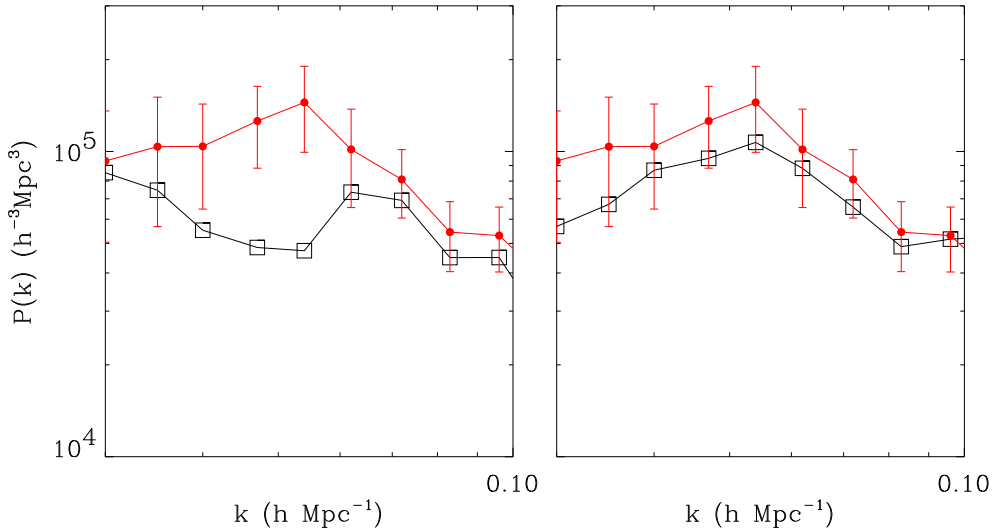}
\caption{Power spectrum of the observed cluster sample (filled
circles) and of two mock cluster samples (squares), extracted from
the same simulation box. As in Fig.~\ref{fig-obs}, 
error bars correspond to the estimated 1$\sigma$
cosmic scatter.}
\label{2mock}
\end{figure}

To illustrate this point we compare in Fig.~\ref{2mock} the power
spectrum of two mock samples which have been extracted from the same
simulation box.  They represent the largest deviation between two mock
samples of one simulation which we have found within the almost 20
simulations made. These mock samples clearly demonstrate the possible
importance of cosmic variance, i.~e.\ the fact that there is only one
observational sample (for one special observer in the Universe). We
conclude that a $P(k)$ as peaked as the observational one does not
represent a very unusual finding.  Therefore, the data may well be
compatible with a power spectrum having a smooth turnover around
$k\simeq 0.05\, h$ Mpc$^{-1}$, as expected for standard scenarios of
structure formation, like those considered here.

On the other hand it is remarkable that Landy et al.\ (1996) also
found a peak at about the same scale from the analysis of the Las
Campanas Redshift Survey. Doroshkevich et al.\ (1997) found a typical
scale of the same order in the Las Campanas Redshift Survey.  Also the
typical diameter of voids in the distribution of clusters is of the
order of this scale (Gottl\"ober et al.\ 1997).  Gazta\~naga and Baugh
(1998) found a steep slope for the real space power spectrum of APM
galaxies in the same range as the unexpectedly steep slope of the
cluster power spectrum.  Eisenstein et al.\ (1998) concluded that an
excess power on a $ 100 \hMpc$ scale can be explained by baryonic
acoustic oscillations only assuming rather extreme regions of the
possible parameter space.  Atrio--Barandela et al.\ (1997) and
Lesgourgues et al.\ (1997) have discussed primordial spectra which
include features that enable these models to account for a possible
excess of power at $\sim 100\,h^{-1}$Mpc scales.

As for the bias parameter, we find that a larger $P(k)$ for the DM
distribution does not correspond in general to a larger $P(k)$ for the
resulting cluster distribution. Indeed, we find a remarkable
amplification for the cluster power spectrum in the case of BSI, with
a cluster biasing factor $b_{\rmn{c}}^2\simeq 15$, while a much
smaller enhancement occurs for SCDM, TCDM, CHDM, $\Lambda$CDM, and
OCDM.  This is consistent with the result that the cluster clustering
does not depend on the amplitude of the underlying DM spectrum, but
only on its shape, once the average number density of clusters is kept
fixed (Croft \& Efstathiou 1994; Borgani et al.\ 1997); the larger the
large--to--small--scale power ratio, the larger the clustering
amplification due to long wavelength modes.  This is the reason why a
model like BSI, which has a large relative amount of power on large
scales and a low normalization, turns out to generate a highly biased
cluster population.

\section{Conclusion}\label{sect:concl}
In this paper we estimated the power spectrum, $P(k)$, for a combined
redshift sample of Abell--ACO clusters. The result is compared with
those obtained for mock cluster samples, which were extracted from
large PM $N$--body simulations of seven different models, so as to
reproduce the selection effects of the Abell--ACO data set. This
analysis allows us not only to discriminate among different models for
cosmic structure formation, but also to understand in detail the
effect of cosmic variance on the power spectrum shape at the largest
accessible scales.

The cluster power spectrum is reliably detected over the scale range
$0.03\la k\la 0.2\,\hMpcI$, over which the analysis of cluster
simulations demonstrates that the window function associated to the
sample geometry has a negligible effect on $P(k)$. For $k\ga
0.05\,\hMpcI$ the cluster power spectrum is well approximated by a
power law, $P(k)\propto k^n$ with $n\simeq -1.9$, while it changes
sharply to a positive slope at smaller wavenumbers. We find a peak in
the power spectrum at $k \simeq 0.05\,\hMpcI$, which, however, is 
not as pronounced as the one detected by Einasto et al.\ (1997a).

The BSI model provides the best fit for the cluster power spectrum of
Abell--ACO clusters among the models that we considered.  In general,
the models fail at about $2\sigma$ level at reproducing a $P(k)$ slope
for $k\ga 0.05\,\hMpcI$ as steep as that of the Abell--ACO sample.
Among the more than 100 mock samples that we extracted from the
simulation boxes for the different models, we found that only one of
them has a power spectrum with a feature almost identical to the
observed one within the range $0.03\la k\la 0.1\,\hMpcI$.  This
finding shows the possible relevance of cosmic variance. It is just
what one would expect for a feature, like the measured peak in the
$P(k)$ shape, which represents a $2 \sigma$ deviation from a smooth
mean spectrum.

Observational samples which were both encompassing larger volumes and
having selection effects under a strict control are required in order
to decide whether details of the clustering pattern on a $100 \hMpc$
scale force us to revise our understanding of structure formation or
just leads to a refinement of the models which are already in play.


\section*{Acknowledgment}
We are grateful to Manolis Plionis who kindly provided the cluster
sample redshift data and for comments on the manuscript.  We are also
indebted to Peter Schuecker for numerous useful conversations and for
remarks on the manuscript. JR acknowledges receipt of grant Go563/5--2
of the Deutsche Forschungsgemeinschaft.  SG acknowledges support from
the Deutsche Akademie der Naturforscher Leopoldina with means of the
Bundesministerium f\"ur Bildung und Forschung grant LPD 1996. AK
acknowledges receipt of NASA ATP grant.


\appendix
\section{The power spectrum estimation}\label{est-tech}

In this Appendix we briefly describe the derivation of the power
spectrum estimator, as given by Eq.~(\ref{final-estimator}).  The
Abell--ACO cluster sample represents a three--dimensional catalogue
which is almost complete (volume--limited) out to its boundaries.  The
selection effects arising from galactic absorption and redshift
extinction are parameterized according to Eqs.\ (1) and (2).  In order
to estimate the power spectrum we employ the standard method, as
described, e.g., by Fisher et al.\ (1993) and Lin et al.\ (1996).

The sample's geometry is described by the window function $W(\vec{r})
\equiv 1$ inside the sample volume and zero otherwise.  Its Fourier
transform has been computed by means of Monte Carlo integration.  A
total of $N=10^5$ random points with position vectors $\vec{r}_i \in
V$ is generated to evaluate the Fourier representation of the window
function $(1/N)\sum_{i}e^{i\vec{k}\cdot\vec{r}_i}$.  The large number
of points guarantees the noise level introduced into the power
spectrum to be more than two decades below the noise level due to the
finite number of galaxy clusters. Therefore, no correction for the
shot noise contribution of the window function is applied.

\begin{figure}
\epsffile{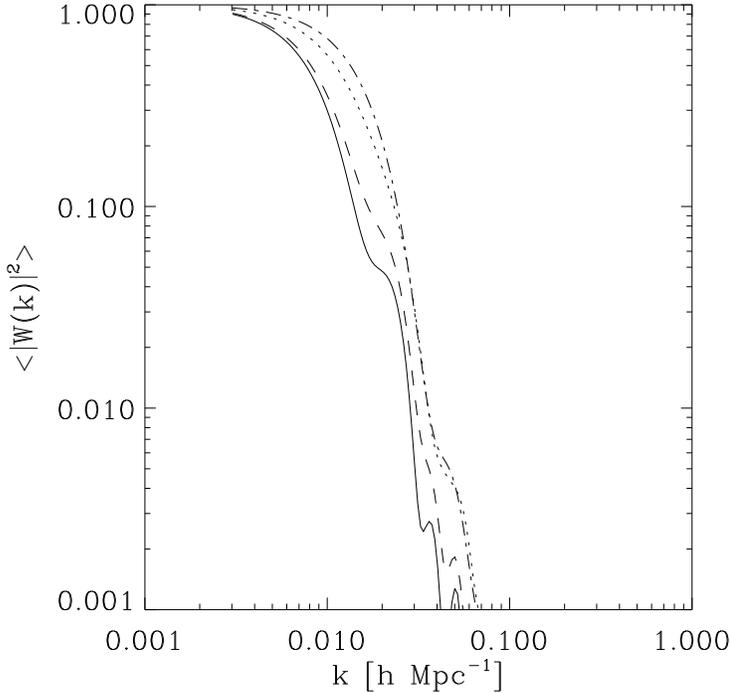}
\caption{
Directional average of the window function (also called window power).
The solid line corresponds to the Abell--ACO sample, the dashed and
dotted lines describe Abell and ACO part separately (southern--galactic
portion of the ACO sample (ACO--SG) --- dash--dotted).}
\label{fig-w2k}
\end{figure}

The double--cone geometry in real space with rotational symmetry about
the $z$--axis (pointing towards the galactic pole) gives rise to a
Fourier transform mostly localized along the $k_z$--axis and with
cylindrical--symmetric side lobes in the $k_x$--$k_y$--plane.  In
Fig.~\ref{fig-w2k} we have plotted the directional average $\langle
|\hat{W}(\vec{k})|^2 \rangle_{|\vec{k}|=k}$ for the combined
Abell--ACO sample (solid line) and for the subsamples Abell (dashed),
ACO (dotted) and ACO--SG (dash--dotted) separately. The maximum
wavelength, whose power can be explored reliably, grows with the
sample's extension.  Note that the depth of the three samples is all
the same but the coverage of the celestial sphere is different. As a
rule of thumb the minimum wavenumber $k_{\rmn{min}}$ is given by
$\langle |\hat{W}(\vec{k})|^2 \rangle_{|\vec{k}|=k_{\rmn{min}}} \equiv
0.1$ (Peacock \& West 1992). This corresponds to wavenumbers
$k_{\rmn{min}}$ of 0.014, 0.017, 0.022 and $0.024\,\hMpcI$ for the
samples Abell--ACO, Abell, ACO and ACO--SG, respectively.

We define the cluster density contrast within the sampled volume as 
\begin{equation}
\delta(\vec{r}) \equiv \frac{1}{n} \sum_{i} 
\frac{\delta^{\rmn{D}}(\vec{r}-\vec{r}_i)}{\phi(\vec{r}_i)}-1\,,
\end{equation}
where $\vec{r}_i$ is the position vector of the $i$--th cluster,
$\phi(\vec{r}_i)$ is the selection function which is used
to model galactic absorption and redshift extinction, $n$ is the mean
cluster number density and $\delta^{\rmn{D}}$ is Dirac's delta
function.  The selection function is computed as the product
\begin{equation} \phi(\vec{r}) = D \varphi(b) \psi(z) \; , \end{equation}
with $D \simeq 0.7$ being the mean density correction applied to Abell
clusters, while $D=1$ for ACO clusters.

The squared modulus of the Fourier transform
$\delta(\vec{k})$ of the density contrast $\delta(\vec{r})$, $\Pi(\vec{k})$, is an
estimate of the power spectrum. The connection between $\Pi(\vec{k})$
and the underlying power spectrum $P(k)$ can be found by calculating
the ensemble mean of $\Pi(\vec{k})$ (see, e.g., Fisher et al.\ 1993
for the details),
\begin{equation} 
\left\langle \Pi(\vec{k}) \right\rangle \equiv \left\langle \left| \delta(\vec{k}) \right| ^2  \right\rangle
= \frac{\tilde{P}(\vec{k})}{V} + S \; ,
\end{equation}
where
\begin{equation}\label{tildeP}
\tilde{P}(\vec{k}) \equiv 
\frac{V}{(2 \pi)^3} \int {\rmn{d}}^3 k' \, P(\vec{k'}) |\hat{W}(\vec{k'}-\vec{k})|^2
\end{equation}
denotes the convolution of the power spectrum with the window function
and
\begin{equation}\label{eq-shotnoise}
S \equiv \frac{1}{nV^2} \int_V {\rmn{d}}^3 r \,
\frac{1}{\phi(\vec{r})} \; .
\end{equation}
the additive shot noise term. The ensemble average must be replaced
with the best estimate $\Pi'(k)$ that can be obtained from the data
set.  The mean number density is estimated from the data set as $ n' =
V^{-1} \sum_{i} \phi(\vec{r}_i)^{-1}$.  Since $n'$ may be in general
different from the true mean number density of clusters, a bias in the
power spectrum estimation is introduced (Peacock \& Nicholson 1991).
The spectrum is systematically underestimated by a factor of
approximately $1-|\hat{W}(k)|^2$.  A correction for this effect, which
becomes relevant at scales approaching the largest scale covered by
the sample, has been applied in Eq.(\ref{conv}).  The shot noise
(Eq.~\ref{eq-shotnoise}) is given by
\begin{equation}\label{eq-est-s}
S' = \frac{1}{(n'V)^2} \sum_{i}\frac{1}{\phi^2(\vec{r}_i)} \; .
\end{equation}
For each mode $k$ we compute the estimate $\Pi'(k)$ by averaging
$|\delta(\vec{k}_j)|^2$ over M = 1000 random directions of the
wavevector $|\vec{k}_j|=k$,
\begin{equation}\label{eq-est-pi}
\Pi'(k) = \frac{1}{M} \sum_{j}\left|\delta(\vec{k}_j)\right|^2, 
\end{equation}
Therefore, the estimator for the convolved power spectrum reads finally
\begin{equation}\label{conv}
\tilde{P}'(k) = \frac{V}{1-|\hat{W}(k)|^2} \left[ \Pi'(k) - S' \right] \; .
\end{equation}
Inserting Eqs.\ (\ref{eq-est-s}) and (\ref{eq-est-pi}) into
Eq.(\ref{conv}) yields 
the convolved power spectrum Eq.~(\ref{final-estimator}).


\section*{References}

Abell G.O., 1958, ApJ, 3, 211

Abell G.O., Corwin H.G., Olowin R.P., 1989, ApJS, 70, 1

Amendola L., Gottl\"ober S., M\"ucket J.P., M\"uller V., 1995, ApJ,
457, 444

Andernach H., Tago E., Stengler--Larrea E., 1995, Astrophys.\ Lett.\ \&
Comm., 31, 27

Atrio--Barandela F., Einasto J., Gottl\"ober S., M\"uller V., Starobinsky
A.A., 1997, JETP Letters, 66, 367

Bahcall N.A., Soneira R.M., 1983, ApJ, 270, 20

Batuski D.J., Bahcall N.A., Olowin R.P., Burns, J.O., 1989, ApJ, 341,
  599

Bardeen J.M., Bond J.R., Kaiser N., Szalay A.S., 1986, ApJ, 304, 15

Bennett C.L. et al., 1994, ApJ, 436, 423

Borgani S., Plionis M., Coles P., Moscardini L., 1995, MNRAS, 277, 1191

Borgani S., Moscardini L., Plionis M., G\'orski K.M., Holtzman J.,
Klypin A., Primack J.P., Smith C.C., 1997, NewA, 1, 321

Cappi A., Maurogordato S., 1992, A\&A, 259, 423

Collins C.A., Guzzo L., Nichol R.C., Lumsden S.L., 1995, MNRAS, 274,
  1071

Croft R.A.C., Efstathiou G., 1994, MNRAS, 267, 390

Croft R.A.C., Dalton G.B., Efstathiou G., Sutherland W.J., Maddox
S.J., 1997, MNRAS, 291, 305

Dalton G.B., Croft R.A.C., Efstathiou G., Sutherland W.J., Maddox
  S.J., Davis M., 1994, MNRAS, 271, L47

Doroshkevich A. G., Tucker D.L., Oemler A. JR., Kirshner R.P., Lin H.,
Shectman S.A., Landy S.D., Fong R., 1996, MNRAS, 283, 1281

Ebeling H., Edge A.C., Fabian A.C., Allen S.W., Crawford
  C.S., 1997, ApJ, 479, L101

Efstathiou G.P., Bond J.R., White S.D.M., 1992a, MNRAS, 258, 1P

Efstathiou G., Dalton G.B., Sutherland W.J., Maddox S.J.,
  MNRAS, 1992b, 257, 125

Einasto J., Gramann M., Saar E., Tago E., 1993, MNRAS, 260, 705

Einasto J., Einasto M., Gottl\"ober S., M\"uller V., Saar V.,
Starobinsky A.A., Tago E., Tucker D., Andernach H., Frisch P., 1997a,
Nature, 385, 139

Einasto J., Einasto M., Frisch P., Gottl\"ober S., M\"uller V., Saar
V., Starobinsky A.A., Tago E., Tucker D., Andernach H., 1997b, MNRAS,
289, 801

Eisenstein D.J., Hu W., Silk J., and Szalay A.S., 1998, ApJ, 494, 1

Feldman H.A., Kaiser N., Peacock J.A., 1994, ApJ, 426, 23

Fisher K.B., Davis M., Strauss M.A., Yahil A., Huchra J.P., 1993, ApJ,
402, 42

Gazta\~naga E., Baugh C.M., 1998, MNRAS, 294, 229

Gazta\~naga E., Croft R.A.C., Dalton G.B., 1995, MNRAS,
  276, 336

Ghigna S., Bonometto S.A., Retzlaff J., Gottl\"ober S., Murante G.,
1996, ApJ, 469, 40

G\'orski K.M., Banday A.J., Bennett C.L., Hinshaw G.,
Kogut A., Smoot G.F., Wright E.L., 1996, ApJ, 464, L11 (G96)

G\'orski K.M., Ratra B. Stompor R., Sugiyama N.,
Banday A.J., 1998, ApJS, 114, 1 (G98)

Gottl\"ober S., 1996, Proceedings of the International School of
Physics ''Enrico Fermi'', eds. S. Bonometto, J.R. Primak,
A. Provenzale, IOS Press, Amsterdam, p. 467

Gottl\"ober S., M\"uller V., Starobinsky A.A., 1991,
Phys. Rev. D43, 2510

Gottl\"ober S., M\"ucket J.P., Starobinsky A.A., 1994,
ApJ, 434, 417

Gottl\"ober S., Retzlaff J., Turchaninov V., 1997, 
Astrophysics Reports 2, 55.

Jenkins A. et al.\ (The Virgo Consortium), 1998, ApJ, 499, 20

Jing Y.P., Plionis M., Valdarnini R., 1992, ApJ, 389, 499

Jing Y.P., Valdarnini R., 1993, ApJ, 406, 6

Kates R., M\"uller V., Gottl\"ober S., M\"ucket J.P., Retzlaff J.,
1995, MNRAS, 277, 1254

Kerscher M., Schmalzing J., Retzlaff J., Borgani S., Buchert T.,
Gottl\"ober S., M\"uller V., Plionis M., Wagner H., 1997, MNRAS, 284,
73

Klypin A., Kopylov A.I., 1983, Soviet Astron. Lett., 9, 41

Klypin A., Rhee G., 1994, ApJ, 428, 399

Klypin A., Gottl\"ober S., Kravtsov A., Khokhlov A., 1997, astro--ph/9708191

Landy S.D., Shectman S.A., Lin H.L., Kirshner R.P., Oemler A., Tucker
D.L., Schechter P.L., 1996, ApJ, 456, L1

Lesgourgues J., Polarski D., Starobinsky A. A., 1998, MNRAS, 297, 769

Lin H.L., Kirshner R.P., Shectman S.A., Landy S.D., Oemler A., Tucker
D.L., Schechter P.L., 1996, ApJ, 471, 617

Lucchin F., Matarrese S., 1985, Phys. Rev., D32, 1316

Nichol R.C., Briel U.G., Henry J.P., 1994, MNRAS, 265, 867

Nichol R.C., Connolly A.J., 1996, MNRAS, 279, 521

Olivier  S.S., Primack J.R., Blumenthal G.R., Dekel A.,
  1993, ApJ, 408, 17

Park C., Vogeley M.S., Geller M.J., Huchra J.P., 1994, ApJ, 431, 569

Peacock J.A., Nicholson D., 1991, MNRAS, 253, 307

Peacock J.A., West M.J., 1992, MNRAS, 259, 494

Plionis M., Valdarnini R., 1991, MNRAS, 249, 46

Plionis M., Valdarnini R., 1995, MNRAS, 272, 869

Plionis M., Valdarnini R., Jing Y.P., 1992, ApJ, 398, 12

Postman M, Huchra J.P., Geller M., 1992, ApJ, 384, 407

Primack J., Holtzman J., Klypin A., Caldwell D., 1995,
Phys. Rev. Lett. 74, 2160

Ratra B., Peebles, P.J.E., 1994, ApJ, 432, L5

Romer A.K., Collins C.A., B\"ohringer H., Cruddace R.C.,
  Ebeling H., MacGillawray H.T., Voges W., 1994, Nature, 372, 75

Scaramella R., Zamorani G., Vettolani G., Chincarini G.,
  1991, AJ, 101, 342

Schuecker P., Ott H.--A., Seitter W.C., 1996, ApJ, 472, 485

Strauss M.A., Willick J.A., 1995, Phys. Rep., 261, 271

Sutherland W.J., 1988, MNRAS, 234, 159

Tadros H., Efstathiou G., 1996, MNRAS, 282, 1381

Tadros H., Efstathiou G., Dalton G., 1998, MNRAS, 296, 995

Viana P.T.P, Liddle A.R., 1996, MNRAS, 281, 323 (V96)

White M., Gelmini G., Silk J., (1995), Phys. Rev. D51, 2669

Wilson M.L., 1983, ApJ, 273, 2.

\end{document}